\documentclass[a4paper,12pt]{article}
\usepackage{graphicx} %use graph format
\usepackage{epstopdf}
\usepackage{ulem}
\usepackage{color}
\usepackage[labelfont=bf,labelsep=space]{caption}

\begin{document}

%%-move to normal A4-%%
\hoffset = -1truecm \voffset = -2truecm \baselineskip = 10 mm

\title{\bf Uncovering underappreciated physical effects hidden in the cosmic-ray electron spectra at very high-energy }

\author{Wei Zhu$^{1,2*}$, Yu-Chen Tang$^3$, Feng-zheng Zhu$^1$
and Bo Yang$^1$
\\\\
         \normalsize 1 Center for Fundamental Physics and School of Mathematics and Physics,\\
         \normalsize Hubei Polytechnic University, Huangshi 435003, P.R. China\\
         \normalsize 2 Department of Physics, East China Normal University, Shanghai 200241, P.R. China\\
         \normalsize 3 Key Laboratory of Dark Matter and Space Astronomy, Purple Mountain Observatory,\\
         \normalsize Chinese Academy of Sciences, Nanjing 210008, P.R. China\\
\normalsize $^*$ Corresponding author, E-mail: wzhu@phy.ecnu.edu.cn.
         }

\newpage

\maketitle

\textbf{Abstract}
We show that the behavior of the cosmic ray electron spectrum in the TeV energy band near the Earth is dominated by gluon condensation and anomalous electron/positron pair-production in Cygnus X.

\textbf{Kerwords}
Cosmic-ray; Gamma-ray; Nuclear Physics; Gluon Condensation

\newpage

\section{Introduction}

    The source of the cosmic-ray electrons (CREs) and the mechanism of their formation can provide important information on the origins of
dark matter, antimatter, and cosmic rays, especially the exploration of new phenomena in the Milky Way [1-5].
Near-Earth observations can directly, or indirectly, record the flow of electrons from space, but there's no way to tell where they're coming from. This makes it difficult to study the origin of CRE. It is a standard tool to interpret the CRE spectrum by theoretically modeling all astrophysical constituents that may contribute to the observed energy spectrum. These components include the sum of the mean electron flow from distant sources and the electron flow from the local supernova remnant in the Green's catalogues [6].
Because of the large degree of freedom in this calculation, it is always possible to find a set of electron sources that fits all the determined CRE spectra. However, in the very high energy region ($E_e>1$ TeV), where there is a lack of observational data, there is a large uncertainty in the predictions of the model due to the number of free parameters involved. We believe that there may be other reasons. Since the estimation of the contribution of individual electron source is based on an indirect extrapolation of the nature of the observed electromagnetic radiation, there is a possibility of misjudgment of the radiation mechanism here, e.g., by treating the hadronic scenario as a leptonic scenario, or even of the existence of a new mechanism that has been overlooked.

    As expected, recently the high energy stereoscopic system (H.E.S.S.) Collaboration has accurately processed data collected over a period
of 12 years, revealing unprecedented detail about the CRE spectrum [7]. Combined with previous measurements of the energy spectrum by AMS, Fermi-LAT, DAMPE and CALET et al., the results show a sharp "break" at $\sim$ 1 TeV, where the slope of the spectrum steepens.
A power law (PL) with little or no curvature correction detected beyond the break lasted a full order of magnitude in energy . This result challenges the origin of the CRE, since any lepton scheme and the conventional hadron scheme predict an exponential decay of the individual high-energy electron spectra in the high-energy region rather than a PL. Although mathematically it is always possible to find a set of curves to fit a section of PL. but this is entirely artificial, not natural selection. It implies that there are possible new physical effects that are not yet clear to us.

    Another recently discovered CRE-related astronomical event is that the LHASSO
recorded a giant vary high energy (VHE) gamma-ray bubble in the star-forming region of Cygnus x, from which a candidate for the origin of cosmic rays has been found, the most probable counterpart of a super cosmic ray source including CRE [8].

    In this article we propose that the gluon condensation (GC) [9] in nucleon may be responsible for the broken power law
(BPL) in the CRE spectrum. Besides, an anomalous electron/positron pair-production [10] in Cygnus x also plays an important role.
The CREs can be generated by the following hadronic scenario. Collisions of energetic protons with interstellar matter (protons and nuclei) produce a large number of pions, of which neutral pions decay into photons, which can be excited in the Coulomb field to produce pairs of positron and electron. Here the cross sections of $pp\rightarrow \pi$ and $\gamma\rightarrow e^+e$
have an important influence on the final CRE spectrum. We find that new insights into these two cross sections can explain the observed CRE spectrum.
We will give the main results for the GC and the improved Bethe-Hitler formula in Sec. 2 and 3 respectively, and then construct a new CRE model in Set. 4. Discussions and conclusion will be given in Sec. 5.

\section{Gluon condensation}

    The pp cross section at high energy ultimately depends on the distribution of gluons in the nucleon.  Measuring the gluon distribution is one
of the tasks of high-energy physics experiments, which is limited by the experimental conditions. Cosmic ray energies far exceed the maximum energies of existing proton accelerators, so there are many predictions of the distribution of gluons in protons based on the QCD theory.
\begin{figure}
	\begin{center}
		\includegraphics[width=0.8\textwidth]{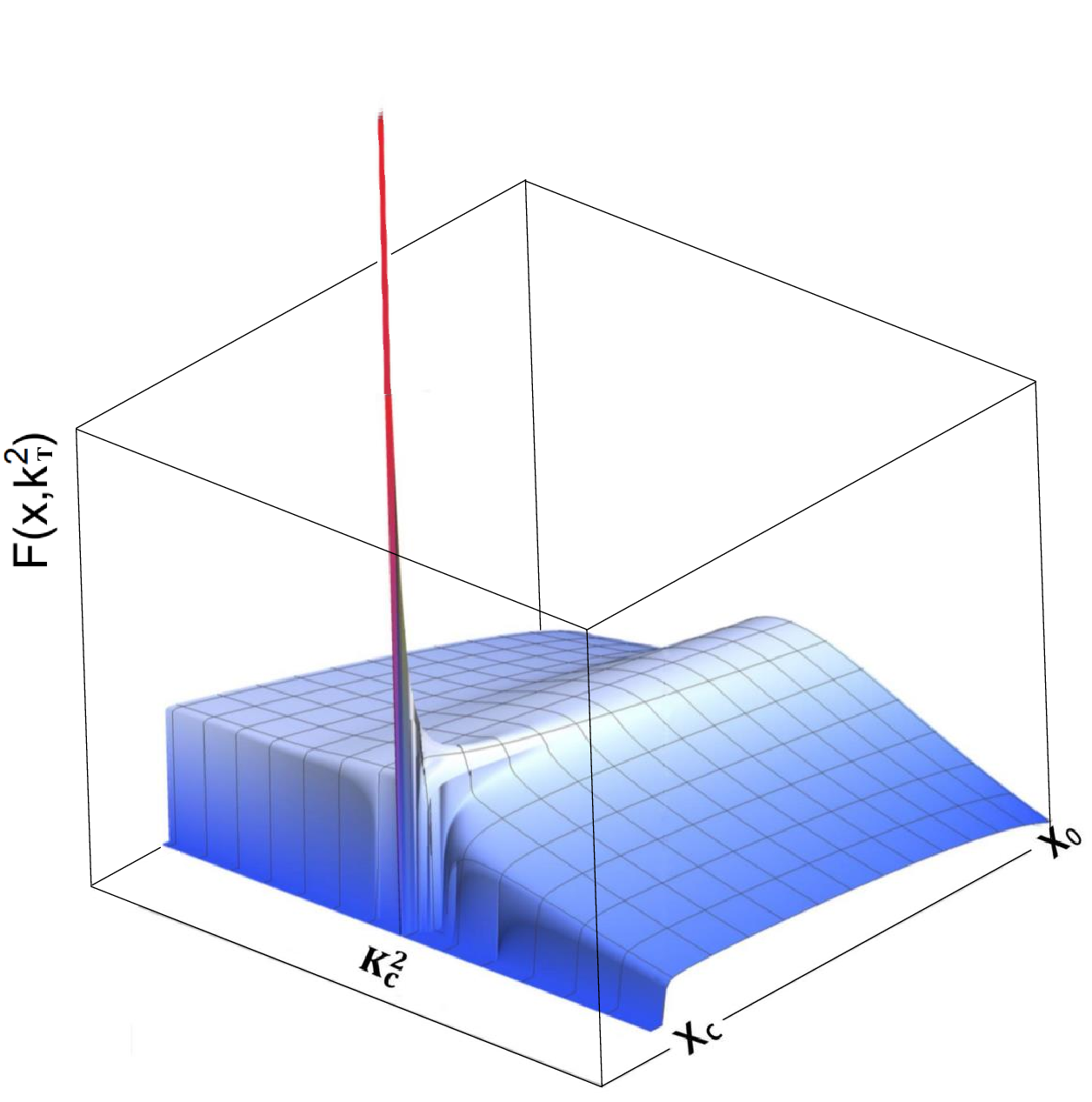}
		\caption{A schematic diagram for the evolution of gluon distribution from the CGC
at $x_0$ to the GC at $(x_c,k^2_c)$ [9].
		}\label{fig:1}
	\end{center}
\end{figure}
One such example is the gluon condensation shown in Fig. 1 [9]. It shows that gluons evolve from the so-called colour glass condensate (CGC) distribution at $(x_c,Q_s^2)$ to smaller $x<x_c$, and condense at a critical point $(x_c,k_c)$, where $x$ is the fraction of the longitudinal momentum of nucleon taken by gluon.
Calculations show that when the centre-of-mass energy of a pp collision reaches $\sqrt{S_c}$, condensed gluons suddenly enter the collision range, generating a large number of pion, which inevitably creates a protrusion in the gamma and electron spectra. Importantly, since gluons are the main  material for making new hadrons (although the exact mechanism is not clear), huge amounts of condensed gluons can saturate the secondary pion production, i.e., almost all of the available kinetic energy of the collision is used exclusively for the manufacture of pions. In this case, using only the two universal conditions of energy conservation and relativistic covariance, it is possible to restrict the gamma energy spectrum to a typical BPL [11],

$$E_{\gamma}^2\Phi^{GC}_{\gamma}(E_{\gamma})=\left\{
\begin{array}{ll}
\frac{2e^bC_{\gamma}}{2\beta_p-1}(E_{\pi}^{GC})^3\left(\frac{E_{\gamma}}{E_{\pi}^{GC}}\right)^{-\beta_{\gamma}+2} & {\rm if~}E_{\gamma}\leq E_{\pi}^{GC}\\\\
\frac{2e^bC_{\gamma}}{2\beta_p-1}(E_{\pi}^{GC})^3\left(\frac{E_{\gamma}}{E_{\pi}^{GC}}\right)^{-\beta_{\gamma}-2\beta_p+3}
& {\rm if~} E_{\gamma}>E_{\pi}^{GC}
\end{array} \right. .\eqno(1)$$in the GeV-unit. This is a typical broken power-law. It would be expected that the corresponding CRE spectrum would show a similar distribution via $\gamma\rightarrow e^+e$. In fact,
these gamma rays may produce electron/positron pair via $\gamma+Coulomb~feld\rightarrow e^+e$ in
the source. The result is [11]

$$\Phi_e^{GC}(E_e)=\left\{
\begin{array}{ll}
\frac{2e^bC_e}{2\beta_p-1}E_{\pi}^{GC}\left(\frac{E_e}{E_{\pi}^{GC}}\right)^{-\beta_e}\left[\frac{1}{\beta_{\gamma}}
\left(\frac{E_e}{E_{\pi}^{GC}}\right)^{-\beta_{\gamma}}+
(\frac{1}{\beta_{\gamma}+2\beta_p-1}-\frac{1}{\beta_{\gamma}})\right]
& {\rm if~} E_e\leq E_{\pi}^{GC}\\\\
\frac{2e^bC_e}{(2\beta_p-1)(\beta_{\gamma}+2\beta_p-1)}
(E_{\pi}^{GC})\left(\frac{
E_e}{E_{\pi}^{GC}}\right)^{-\beta_e-\beta_{\gamma}-2\beta_p+1} &
{\rm if~} E_e>E_{\pi}^{GC}
\end{array} \right. \eqno(2)$$

    However, the probability of the above process producing the electron/positron-pair is very small. It is generally believed that
high-energy electrons are formed directly from the acceleration of the original low-energy electrons. However, the following advance about the Bethe-Heitler formula offers a possibility of generating high-energy CREs from a hadronic scenario.

\section{Improved Bethe-Heitler formula}

    The interaction between charged particles and matter is a central theme in the study of high-energy phenomena in the universe.
The Bethe-Heitler formulae for bremsstrahlung and electron/positron pair-production are widely used to study high-energy processes in a wide variety of celestial systems.
Work [9] indicates that if the recoil of the target can be ignored in a certain high-energy range,
both the electron/positron pair-production and bremsstrahlung cross sections have an unexpectedly large increment, which is ignored by the Bethe-Heitler formula, i.e., the Bethe-Heitler formula

$$d\sigma_{\gamma\rightarrow e}=\frac{\alpha^3}{m_e^2}\ln\frac{4\omega^2}{\mu^2}(1-z)
[(1-z)^2+z^2]dz, \eqno(3) $$becoms a modified Bethe-Heitler formula

$$d\sigma_{\gamma\rightarrow e}=\frac{\alpha^3}{\mu^2}\ln\frac{4\omega^2}{\mu^2}(1-z)
[(1-z)^2+z^2]dz. \eqno(4) $$ One can find a large increase in the cross section due to the replacement of $m_e$ by $\mu$ in the denominator,
where the screening parameter $\mu$ is several orders of magnitude smaller than $m_e$. Thus, there is an energy window, in which the pair-production cross section is anomaly enhanced. In dilute ionised gases, the cross section can be enhanced by more than eight orders of magnitude [12].
In this way, based on the GC model, we give a self-consistent interpretation of the CER spectrum and its origin in the following two Sections.

\section{The CRE spectrum}

    As we know that in addition to a strong gamma-electron source close to the Earth, the observational
data also include electron contributions from other sources. The Milky Way has many electron sources, they can be pulsars, supernova remnants
(SNRs),  and active galactic nuclei (AGNs), even gamma-ray bursts (GRBs). They are large in number, widely distributed and have randomly changing electron orbits. We include them all in the background contribution. This background correction has a significant impact on the interpretation of the CRE spectrum, and the true physical background is generally simulated with a diffusion model. This physical background
is not smooth, has undulations, depending on the specific model. However, when the contribution of a source is particularly strong (for example, Cygnus), the corrections of the background fluctuations are relatively weak and negligible. Thus, the empirical PL is a good approximation to depict the background, but such an extension will inevitably go beyond the broken spectrum after 1 TeV, and appear unreasonably negative.
negative values, so an exponential truncation factor is added. A commonly used empirical formula for the background is

$$E_e^3\Phi_e^0(E_e)=C E_e^{-\beta}\exp[-(E_e/E_c)^{\delta}], \eqno(5)$$$\delta$ regulating the rate of decline. The possible reasons for this cut phenomenon are that the high-energy end of the background electron flow may be limited by the acceleration source or high-energy electrons are prone to escape from the Milky Way. A reasonable conjecture is the existence of a strong cosmic ray source in the vicinity of the Earth (relative to galactic scale) whose gamma spectrum near 1 TeV exhibits the BPL. We find Cygnus to be a possible late pick. Cygnus x is one of the densest and most neighbourly (at a distance of $\sim 4.5$ kpc) star-forming region in the Milky Way. It contains enormous H and molecular gas. It also contains cosmic ray accelerators.

    Recently LHASSO recorded the gamma-ray spectrum from Cygnus x, which exhibits a typical PL above 1 TeV with the
photon index of 2.71 [10]. It corresponds to $\beta_{\gamma}+2\beta_p=3.8$ in Eq. (1). In order to reduce the uncertainty in the contribution of the electronic background, we add this condition when fitting the CRE spectrum.

    Using the minimum-squares fitting procedure, the CRE spectrum was fitted with the DAMP data at $E_e<E_{\pi}^{GC}$ and the H.E.S.S. data at
$E_e>E_{\pi}^{GC}$ with eight free parameters.  The electron background was fitted by the AMS data at the low-energy.
The result is shown in Fig. 2. The fitting goodness is an ideal value $\chi^2/dof= 0.37$.
As shown in Eq. 2b, the right side is a typical PL, while the left side has the curvature correction. The background decreases exponentially at $E_e>1$ TeV and gives rise to a GC-specific PL.
Note that all leptonic scenarios and conventional hadronic models predict photon- and electron-spectra that are exponentially decreasing at $E>1$ TeV, only the GC model has a very broad PL form at high-energy, and they have been supported by numerous astronomical events [13-15].

\begin{figure}
	\begin{center}
		\includegraphics[width=0.8\textwidth]{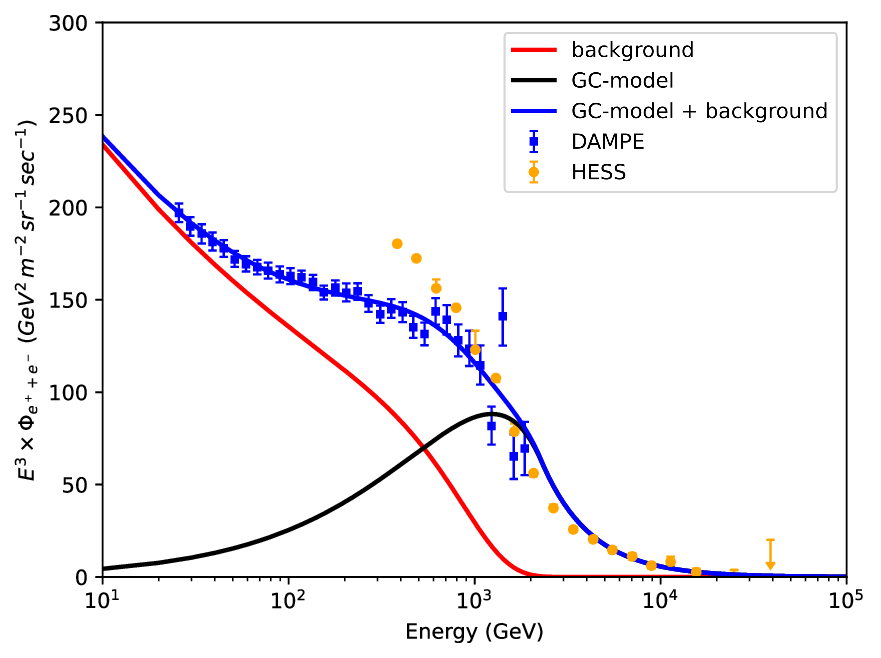}
        \renewcommand{\thefigure}{2a}
		\caption{The CRE spectrum (multiplied by $E_e^3$ predicted by the GC model Eq. (2) and comparison with the data measured by DAMPE [3]
and H.E.S.S. [7]. The parameters for background are $C=400~GeV^{2.233}m^{-2}sr^{-1}sec^{-1}$, $\beta=0.233$, $E_c=1$ TeV and $\delta=2.0$; for GC spectrum, we have $C_e=1.55E-13\pm2.37E-13~ GeV^{-2} m^{-2} sr^{-1} sec^{-1}$, $\beta_{\gamma}=3.796-2\beta_p=0.31$, $\beta_p=1.74\pm0.40$, $\beta_e=1.83\pm0.54$ and $E_{pi}^{GC}=2.29\pm0.27$ TeV.
}\label{fig:2a}
	\end{center}
\end{figure}

\begin{figure}
	\begin{center}
		\includegraphics[width=0.8\textwidth]{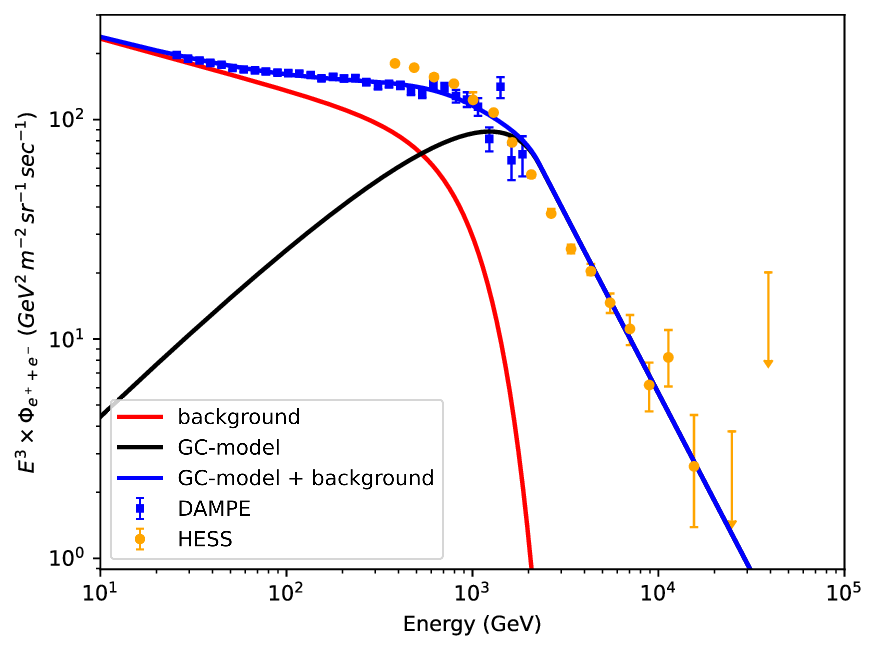}
        \renewcommand{\thefigure}{2b}
		\caption{Same as fig. 2a but with longitudinal logarithmic scale.
		}\label{fig:2b}
	\end{center}
\end{figure}

    Cygnus X has an accelerated cosmic ray cocoon, rich in electrons, protons and ionised light nuclei, where multiple
radiation regimes coexist.  The distribution of TeV energy bands observed by LHASSO has a typical PL, which can be interpreted as being dominated by the GC model. Because the above-mentioned two spectra originate from the same proton acceleration mechanism, we keep $\beta_p=1.74$. On the other hand, since gamma loss occurs within the same Cygnus, we fix $\beta_{\gamma}=0.31$. Substituting them to Eq. (1), we obtain  Fig. 3.
One can find that the GC model predicts a good fit of the PL to the LHASSO observations. Note that according to the GC model,
this PL distribution can extend up to $\sim 14(E_{\pi}^{GC}/GeV)^2\sim 50~PeV$ if the protons are accelerated to a high enough energy region, [11].

\begin{figure}
	\begin{center}
		\includegraphics[width=0.8\textwidth]{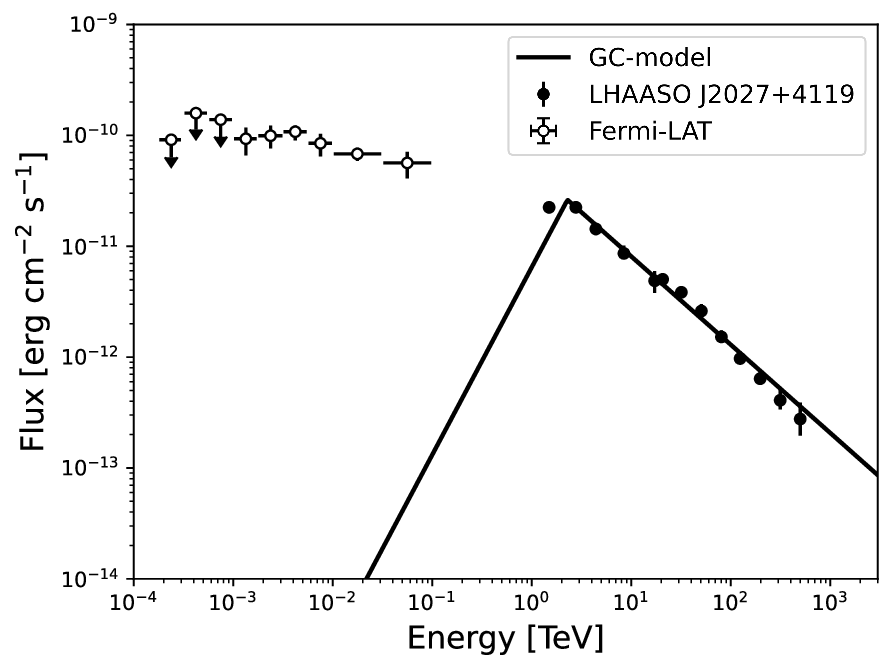}
        \renewcommand{\thefigure}{3}
		\caption{Predicted gamma-ray spectrum from Cygnus by the GC model Eq. (1), where $C_{\gamma}=3.31E-14\pm 1.24E-15~TeV^{-2} cm^{-2} s^{-1}$,
 but $E_{\pi}^{GC}=2.29$ TeV, $\beta_{\gamma}=0.31$ and $\beta_p=1.74$ as same as the CRE spectrum. Data are taken from
 [8,12].}\label{fig:3}
	\end{center}
\end{figure}

    The Fermi-LAT has also recorded a variety of GeV gamma spectra [16]. Interestingly they are stronger than the GC spectrum. The
gamma signals produced by the lepton scenario are shown to mask the weaker GC spectra. However, a question arises as to why the electron signal produced by the former does not appear above the electron spectrum ? One qualitative explanation is that the GeV gamma recorded by FermiLAT is produced by the lepton scenario.
High-energy photons create electron/positron pairs in the electro's Coulomb field.  Due to the small mass of electron, there is a strong recoil. The cross section is dominated by Eq. (3). While the GC spectrum produces electrons/postron pair in the proton Coulomb field. When the impact parameter is far from the proton, the recoil is negligible. The corresponding cross section is Eq.(4). The above difference explains why the FermiLAT GeV photon contribution is not shown in the CRE spectrum of Fig. 2.

\section{Discussions and summary}

(i) Why are there several VHF gamma sources in the Milky Way, but only Cygnus electrons have a significant effect on the electron energy spectrum reaching the Earth?  One of the reasons is that Cygnus is the closest counterpart in the Milky Way that contains a super cosmic ray source, and it continuously produces an abundance of VHE protons, inducing a GC effect that generates powerful gamma rays. The electric field of the ionised gas in Cygnus converts some of these gamma rays into electrons and positrons according to the improved Bethe-Heitler formula. The lifetime and propagation distance of electrons is severely limited by the energy loss from synchrotron radiation .In the standard
cosmic ray transport model this means that the TeV electron source must be local (distance ~1 kpc). However, anomalous electron pair-birth effects can effectively increase the strength of the Cygnus electrons, causing them to affect the observed CRE spectrum on the Earth.

(ii) How to deal with the contribution of near-Earth point sources that have been used in the standard tool? Firstly, we point out that the intensity of the electron flow in the vicinity of the source, obtained on the basis of electron radiation, and from which the intensity of these electrons reaching the Earth is deduced, is subject to large uncertainties, in particular with respect to the attenuation factor for electrons $>$1 TeV. This is evident from the different predictions of many such models. In addition, although the Cygnus bubble is located about 4.5 kpc away. The improved Bethe-Heitler  effect can increase the $\gamma\rightarrow ee^+$ conversion rate by many orders of magnitude.  A more important reason is that primordial electron acceleration in the lepton scenario is suppressed from 1 TeV due to radiation loss. Whereas in the hadronic scenario with the GC effect the primordial proton does not start to be suppressed until it accelerates to a higher knee, so that the next $\gamma\rightarrow ee^+$ produced electrons can remain in a hard PL form. Therefore, we assume that the electron flow from Cygnus dominates the high-energy behaviour of the CRE, while contributions from other near-Earth sources are classified as background flows.

(iii) Why does the GC model break at 2.3 TeV? The break $E_{\pi}^{GC}$ corresponds to the start of the GC peak into the pA collisions. The QCD evolution equation predicts that $E_{\pi}^{GC}$ changes with the target nucleus. A series of GC spectral analyses revealed that $E_{\pi}^{GC}=20TeV-1TeV$ for light nuclei [17]. We estimate that a molecular assemblage of 2.3 TeV dominates the process.

(iv) The AMS mass spectrometer can measure the positron spectrum. The intensity of $E_{e^+}^3\Phi_{e^+}$ at $\sim 400$ GeV is only 2/15 of that of an electron. This means that the positron is lost due to annihilation with electrons during propagation from the source to the Earth. Therefore, we ignore its contribution. We noticed that the H.E.S.S. data show that 400-500 GeV has three points above our predicted spectrum. It is possible that this is the neglected positron contribution.

(v) A large number of pions with a certain energy accumulate in a narrow space during each collision. Due to the overlap of their wave functions, they may transform into each other during the formation time, i.e., $\pi^+ \pi^- \rightleftharpoons 2\pi^0$. However, since $m_{\pi^+} + m_{\pi^-} > 2m_{\pi^0}$ and the lifetime of $\pi^0$ ($10^{-16}$ s) is much shorter than the typical weak decay lifetime of $\pi^{\pm}$ ($10^{-6}$ s - $10^{-8}$ s), the equilibrium will be broken, and $\pi^0$ will dominate the secondary processes, allowing us to neglect the contribution of $\pi^{\pm}$.
Of course we can't rule out that there are still small amounts of $\pi^{\pm}$ in the Cygnus source, which emit weak neutrino streams. A study based on Ice Cube data reports that neutrino events are detected in the direction of the Cygnus bubble at a level of about $3\sigma$ above the background noise [18]. Although this does not yet meet the criteria for discovery, it implies that there are weak, suppressed neutrino streams, rather than the strong neutrino streams predicted by traditional hadronic scenario.

(vi) The BPL was first described in energy spectrum of all particles of primary cosmic rays, which spans ten orders of magnitude.
The diffusion models consider the PL to be a statistical average of the contributions from a large number of sources, while the breaks are related to the large scale of the universe. For example, the bulk of cosmic rays up to at least an energy of 1 PeV is believed to originate from within our galaxy. Above that energy which is associated with the so-called "knee," the spectrum steepens.
Although the BPL also occurs in some small-scale (1-2 orders of magnitude) range of gamma ray spectra, and the break points float,
this has been documented in many astronomical observations, but we have not yet recognized it. AS we know that
any single leptonic scenario and the traditional hadronic scenario predict that the energy spectra are exponentially decaying, i.e., there is a curvature-corrected approximation to the BPL. The only exception to this is the GC model, which has a deep physical basis in GC. As demonstrated in this paper, it is not difficult to distinguish between them using high-precision spectroscopic measurements

     In summary, we propose the following picture of the CRE spectrum. Protons are accelerated to unprecedentedly high energies in Cygnus,
colliding with the abundant light-nucleated gas in the region to produce $\pi^0\rightarrow\gamma\rightarrow e^+e$. Due to the large number of condensed gluons entering the $\pi$-production process, the gamma-energy spectrum produced by the above cascade takes on the BPL form, which dominates the electron/positron spectrum. At the same time, the anomalous electron/positron pair-production effect enhances flow intensity of these electrons, forming a characteristic BPL spectrum on the background of the electron flow.

\newpage

\end{document}